\documentclass[12pt]{article}
\usepackage{amssymb,amsfonts}
\usepackage{epsf,epsfig}
\newcounter{map}
\newcounter{fel}
\newcounter{aw}
\newcounter{aww}
\begin{document}
\baselineskip 18pt

\title{
Origin of the second coherent peak in the dynamical structure factor of an asymmetric spin-ladder
}
\author{P.~N.~Bibikov and M.~I.~Vyazovsky\\ \it V.~A.~Fock Institute of Physics,
Sankt-Petersburg State University}

\maketitle

\vskip5mm

\begin{abstract}
Appearance of the second coherent peak in the dynamical structure factor of an asymmetric spin
ladder is suggested. The general arguments are confirmed by the
first order (with respect to the asymmetry) calculation  for a spin ladder with singlet-rung ground
state. Basing on this result a new interpretation is proposed for the inelastic neutron scattering
data in the spin gap compound CuHpCl for which some of the corresponding interaction constants are
estimated.
\end{abstract}

\section{Introduction}
Inelastic neutron scattering is an effective method for analysis of low-energy excitations
in low-dimensional spin systems \cite{1}. The dynamical structure factor (DSF)
obtained from this experiment produces an essential information about the low-energy spectrum.
Sharp peaks of the DSF line shape correspond to coherent modes while broad bands
originate from incoherent excitation continuums.

Theoretical study of a spin ladder DSF was developed in the papers \cite{2},\cite{3}. A strong
antiferromagnetic rung coupling corresponds to the DSF with a single coherent-mode peak \cite{2},
while for a weak coupling the line shape has only an incoherent background \cite{3}.
The models studied in \cite{2},\cite{3} are symmetric under exchange of the legs because
their couplings along both legs are equal to each other and the same is true for the couplings
along both diagonals. Such requirement fails for an asymmetric spin ladder.

The compound ${\rm Cu}_2{\rm (C}_5{\rm H}_{12}{\rm N}_2{\rm )}_2{\rm Cl}_4$ (CuHpCl) was first
interpreted as an asymmetric spin ladder \cite{4} (with non equal couplings along diagonals).
However neutron scattering \cite{5},\cite{6} revealed
two coherent peaks in the DSF line shape for CuHpCl. Since this type of behavior does not agree
with the results of \cite{2}, \cite{3} (obtained for the symmetric case!) it was suggested in
\cite{6} that the magnetic structure of CuHpCl is inconsistent with the spin ladder model.

In this paper we show the principal difference between excitation spectrums of symmetric and
asymmetric spin ladders and
suggest argumentation confirming the existence of the second coherent peak in DSF of a
asymmetric spin ladder. As an example we calculate the DSF for a weakly asymmetric spin ladder with
singlet-rung ground state and produce an evidence for the second coherent peak.

\section{Hamiltonian for an asymmetric spin ladder}
The general Hamiltonian of an asymmetric spin-ladder has the following form,
\begin{equation}
{\hat H}={\hat H}^{symm}+{\hat H}^{asymm},
\end{equation}
where
${\hat H}^{symm}=\sum_nH^{symm}_{n,n+1}$ and $\hat H^{asymm}=\sum_nH^{asymm}_{n,n+1}$.
The local Hamiltonian densities are given by the following expressions:
\begin{equation}
H^{symm}_{n,n+1}=H^{rung}_{n,n+1}+H^{leg}_{n,n+1}+H^{frust}_{n,n+1}+H^{cyc}_{n,n+1}.
\end{equation}
where
\begin{eqnarray}
H^{rung}_{n,n+1}&=&J_{\bot}{\bf S}_{1,n}\cdot{\bf
S}_{2,n},\nonumber\\
H^{leg}_{n,n+1}&=&J^{symm}_{\|}({\bf S}_{1,n}\cdot{\bf S}_{1,n+1}+{\bf
S}_{2,n}\cdot{\bf
S}_{2,n+1}),\nonumber\\
H^{frust}_{n,n+1}&=&J^{symm}_{frust}({\bf S}_{1,n}\cdot{\bf
S}_{2,n+1}+{\bf S}_{2,n}\cdot{\bf
S}_{1,n+1}),\nonumber\\
H^{cyc}_{n,n+1}&=&J_{c}(({\bf S}_{1,n}\cdot{\bf S}_{1,n+1})({\bf
S}_{2,n}\cdot{\bf S}_{2,n+1})+({\bf S}_{1,n}\cdot{\bf
S}_{2,n})({\bf S}_{1,n+1}\cdot{\bf S}_{2,n+1})\nonumber\\
&-&({\bf S}_{1,n}\cdot{\bf S}_{2,n+1})({\bf S}_{2,n}\cdot{\bf
S}_{1,n+1})),
\end{eqnarray}
and
\begin{equation}
H^{asymm}_{n,n+1}=J^{asymm}_{\|}({\bf S}_{1,n}\cdot{\bf
S}_{1,n+1}-{\bf S}_{2,n}\cdot{\bf
S}_{2,n+1})+J^{asymm}_{frust}({\bf S}_{1,n}\cdot{\bf
S}_{2,n+1}-{\bf S}_{2,n}\cdot{\bf S}_{1,n+1}).
\end{equation}
This structure is schematically represented on the Fig.1.
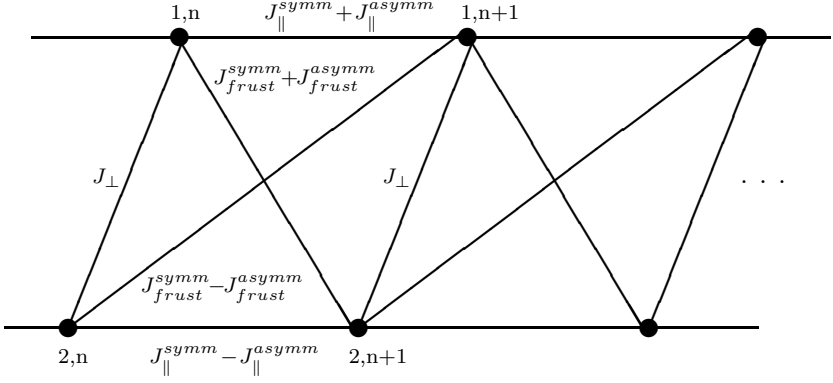
\begin{figure}
\setlength{\unitlength}{3000sp}
\begingroup\makeatletter\ifx\SetFigFont\undefined
\gdef\SetFigFont#1#2#3#4#5{
  \reset@font\fontsize{#1}{#2pt}
  \fontfamily{#3}\fontseries{#4}\fontshape{#5}
  \selectfont}
\fi\endgroup

\begin{picture}(6891,3034)(43,-3110)
{
{\thinlines
\put(601,-2761){\circle*{150}}
\put(3001,-2761){\circle*{150}}
\put(5401,-2761){\circle*{150}}
\put(1521,-361){\circle*{150}}
\put(3901,-361){\circle*{150}}
\put(6301,-361){\circle*{150}}
}{
\thicklines
\put(301,-361){\line( 1, 0){900}}
\put(1201,-361){\line( 1, 0){2400}}
\put(3601,-361){\line( 1, 0){2400}}
\put(6001,-361){\line( 1, 0){900}}
\put( 76,-2761){\line( 1, 0){525}}
\put(601,-2761){\line( 1, 0){2400}}
\put(3001,-2761){\line( 1, 0){2400}}
\put(5401,-2761){\line( 1, 0){900}}
}{
\thicklines
\put(601,-2761){\line( 2, 5){951.724}}
\put(3001,-2761){\line( 2, 5){951.724}}
\put(5401,-2761){\line( 2, 5){951.724}}
\put(601,-2761){\line( 4, 3){3240}}
\put(3001,-2761){\line( 4, 3){3240}}
\put(1511,-361){\line( 3,-5){1456}}
\put(3911,-361){\line( 3,-5){1456}}
}
\put(1901,-211){\makebox(0,0)[lb]
{\smash{{\SetFigFont{8}{14.4}{\rmdefault}{\mddefault}{\updefault}
{$J^{symm}_{\|} \!+\! J^{asymm}_{\|}$}}}}}
\put(950,-3055){\makebox(0,0)[lb]
{\smash{{\SetFigFont{8}{16.8}{\rmdefault}{\mddefault}{\updefault}
{$J^{symm}_{\|} \!-\! J^{asymm}_{\|}$}}}}}
\put(480,-1561){\makebox(0,0)[lb]
{\smash{{\SetFigFont{8}{14.4}{\rmdefault}{\mddefault}{\updefault}
{$J_{\bot}$}}}}}
\put(2880,-1561){\makebox(0,0)[lb]
{\smash{{\SetFigFont{8}{14.4}{\rmdefault}{\mddefault}{\updefault}
{$J_{\bot}$}}}}}
\put(1480,-740){\makebox(0,0)[lb]
{\smash{{\SetFigFont{8}{14.4}{\rmdefault}{\mddefault}{\updefault}
{$J^{symm}_{frust} \!\!+\!\! J^{asymm}_{frust}$}}}}}
\put(880,-2465){\makebox(0,0)[lb]
{\smash{{\SetFigFont{8}{14.4}{\rmdefault}{\mddefault}{\updefault}
{$J^{symm}_{frust} \!\!-\!\! J^{asymm}_{frust}$}}}}}
\put(1126,-211){\makebox(0,0)[lb]
{\smash{{\SetFigFont{8}{14.4}{\rmdefault}{\mddefault}{\updefault}
{1,n}}}}}
\put(3526,-211){\makebox(0,0)[lb]
{\smash{{\SetFigFont{8}{14.4}{\rmdefault}{\mddefault}{\updefault}
{1,n+1}}}}}
\put(200,-3055){\makebox(0,0)[lb]
{\smash{{\SetFigFont{8}{14.4}{\rmdefault}{\mddefault}{\updefault}
{2,n}}}}}
\put(2600,-3055){\makebox(0,0)[lb]
{\smash{{\SetFigFont{8}{14.4}{\rmdefault}{\mddefault}{\updefault}
{2,n+1}}}}}
\put(5826,-1561){\makebox(0,0)[lb]
{\smash{{\SetFigFont{10}{14.4}{\rmdefault}{\mddefault}{\updefault}
{. . .}}}}}
}
\end{picture}
\caption{Schematic of the magnetic structure of an asymmetric spin ladder.}
\end{figure}

It is convenient to extract from the general $\hat H^{symm}$ the "singlet-rung" part $\hat H^{s-r}$
commuting with the following operator:
\begin{equation}
\hat Q=\sum_n Q_n,
\end{equation}
where $Q_n=\frac{1}{2}({\bf S}_{1,n}+{\bf S}_{2,n})^2$. The commutativity condition
\begin{equation}
[\hat H^{s-r},\hat Q]=0,
\end{equation}
or in equivalent form
\begin{equation}
{[}H^{s-r}_{n,n+1},Q_n+Q_{n+1}{]}=0,
\end{equation}
results to the following restriction on the interaction constants for $\hat H^{s-r}$
\begin{equation}
J_{frust}^{s-r}=J_{\|}^{s-r}-\frac{1}{2}J_c^{s-r}.
\end{equation}
According to (8) the Hilbert space for
${\hat H}^{s-r}$ splits on the infinite set of eigenspaces corresponding to different
eigenvalues of $\hat Q$ \cite{7},\cite{8}:
\begin{equation}
{\cal H}=\sum_{m=0}^{\infty}{\cal H}^m,\quad \hat Q|_{{\cal H}^m}=m.
\end{equation}
The one-dimensional subspace ${\cal H}^0$ is generated by the single vector
\begin{equation}
|0\rangle=\prod_n|0\rangle_n,
\end{equation}
where $|0\rangle_n$ is the $n$-th rung singlet. The following restrictions:
\begin{equation}
J^{s-r}_{\bot}>2J^{s-r}_{\|},\quad
J^{s-r}_{\bot}>\frac{5}{2}J^{s-r}_c,\quad J^{s-r}_{\bot}+J^{s-r}_{||}>\frac{3}{4}J^{s-r}_c,
\end{equation}
guarantee that the state (10) is the (singlet-rung) ground state for ${\hat H}^{s-r}$.
The operator $\hat Q$ has a sense of the magnon number \cite{8} associated with $\hat H^{s-r}$.

Decomposition
\begin{equation}
{\hat H}^{symm}={\hat H}^{s-r}+\Delta{\hat H}^{symm}.
\end{equation}

will be correct only if we put some additional restrictions
on the interaction constants of $\Delta\hat H^{symm}$. Postulating them in the form
\begin{equation}
\Delta J_{\bot}^{symm}=\Delta J_c^{symm}=0,\quad \Delta J_{\|}^{symm}=-\Delta J_{frust}^{symm}.
\end{equation}
we guarantee the uniqueness of the decomposition (12).
Moreover under (13) and (4) the local exchange relations between $\Delta\hat H^{symm}$,
$\hat H^{asymm}$ and $\hat Q$ have the following form:
\begin{eqnarray}
\{\Delta H_{n,n+1}^{symm},Q_n+Q_{n+1}\}&=&2\Delta H^{symm}_{n,n+1},\\
\{H^{asymm}_{n,n+1},Q_n+Q_{n+1}\}&=&3H_{n,n+1}^{asymm},
\end{eqnarray}
(where $\{\,,\,\}$ means anti commutator).

As it follows from (14) the term $\Delta\hat H^{symm}$ does not mix even and odd components in
(9). Therefore the Hilbert space ${\cal H}$ splits on two invariant subspaces, of the operator
${\hat H}^{symm}$
\begin{equation}
{\cal H}={\cal H}^{even}+{\cal H}^{odd},\quad {\cal H}^{even}=\sum_{m=0}^{\infty}{\cal H}^{2m},
\quad {\cal H}^{odd}=\sum_{m=0}^{\infty}{\cal H}^{2m+1}.
\end{equation}

From (15) follows that ${\hat H}^{asymm}$ mixes ${\cal H}^{even}$ and ${\cal H}^{odd}$,
however on the sector ${\cal H}^0$ its action is trivial. Really according to (15)
$H_{n,n+1}^{asymm}|0\rangle_n|0\rangle_{n+1}$ have to lie in the sector with $Q_n+Q_{n+1}=3$ that
is impossible because the operator $Q_n$ has only eigenvalues 0 and 1. So we have
\begin{equation}
{\hat H}^{asymm}|0\rangle=0.
\end{equation}
More detailed analysis of the $16\times16$ matrix $H^{asymm}$ (which represent the action of
$H_{n,n+1}^{asymm}$ on the product of $n$-th and $n+1$-rungs) shows that it has only three
(degenerative) eigenvalues: 0, and $\pm\sqrt{(J_{\|}^{asymm})^2+J_{frust}^{asymm})^2}$.
Therefore for small
$J_{\|}^{asymm}$ and $J_{frust}^{asymm}$ the state
$|0\rangle$ remains to be the ground state for $\hat H^{s-r}+\hat H^{asymm}$.

Now we may suggest the following interpretation for the appearance of the second coherent mode in
the DSF line shape of an asymmetric spin ladder. It is known \cite{7}-\cite{9} that in the strong
rung-coupling regime an
excitation spectrum of a symmetric spin ladder has coherent modes of two types,
the one-magnon triplet state lying in ${\cal H}^{odd}$ and three bound two-magnon
states (with total spin 0,1,2) lying in ${\cal H}^{even}$. The ground state also lies in
${\cal H}^{even}$.
In the Born approximation a scattering neutron creates a new state by flipping a single
elementary spin. It is a principal fact that the excited state lies in ${\cal H}^{odd}$.
By this reason in the symmetric case only the subspace
${\cal H}^{even}$ excites during the the scattering process.
However even a little asymmetry results to excitations from ${\cal H}^{even}$ and in particular the
bound two-magnon mode with total spin 1 which is respective for the appearance of the second
coherent peak in the DSF.

In the next sections we shall confirm our arguments by studying the simplest model for which
$\Delta{\hat H}^{symm}=0$ and the ground state exactly has the form (10).

\section{One and two-magnon states for ${\hat H}^{s-r}$}

The eigenstates of $\hat H^{s-r}$ in
the sectors with $\hat Q=1$ and $\hat Q=2$ may be obtained exactly \cite{7},\cite{8}.
From now we shall concern on this special model omitting the upper indexes $"s-r"$ or $"symm"$
in notation of interaction constant $J_c$, $J_{\|}$ and $J_{frust}$. In other words we shall study
the model (2)-(3) with additional restrictions (8) and (11) on $J_{\bot}$, $J_{\|}$, $J_{frust}$
and $J_c$.

According to the following formulas:
\begin{eqnarray}
H_{n,n+1}^{s-r}|0\rangle_n|1\rangle_{n+1}^{\alpha}&=&
(\frac{1}{2}J_{\bot}-\frac{3}{4}J_c)|0\rangle_n|1\rangle_{n+1}^{\alpha}+
\frac{J_c}{2}|1\rangle_n^{\alpha}|0\rangle_{n+1},\nonumber\\
H_{n,n+1}^{s-r}|1\rangle_n^{\alpha}|0\rangle_{n+1}&=&
(\frac{1}{2}J_{\bot}-\frac{3}{4}J_c)|1\rangle_n^{\alpha}|0\rangle_{n+1}+
\frac{J_c}{2}|0\rangle_n|1\rangle_{n+1}^{\alpha},\\
H_{n,n+1}^{s-r}\varepsilon_{\alpha\beta\gamma}|1\rangle_n^{\beta}|1\rangle_{n+1}^{\gamma}&=&
(J_{\bot}-J_{\|}-J_c/4)\varepsilon_{\alpha\beta\gamma}|1\rangle_n^{\beta}|1\rangle_{n+1}^{\gamma},
\end{eqnarray}
where $\alpha,\beta,\gamma=1,2,3$ and $|1\rangle_n^{\alpha}$, is the triplet associated with
a $n$-th rung:
\begin{equation}
|1\rangle_n^{\alpha}=({\bf S}^{\alpha}_{1,n}-{\bf S}^{\alpha}_{2,n})|0\rangle,\quad
({\bf S}^{\alpha}_{1,n}+{\bf
S}^{\alpha}_{2,n})|1\rangle^{\beta}_n=i\varepsilon_{\alpha\beta\gamma}|1\rangle^{\gamma}_n,
\end{equation}
one- and (spin-1) two-magnon states for ${\hat H}^{s-r}$ have the
following form \cite{7},\cite{8}:
\begin{eqnarray}
|k, magn\rangle^{\alpha}_{0}&=&
\frac{1}{\sqrt{N}}\sum_ne^{ikn}...|0\rangle_{n-1}|1\rangle_n^{\alpha}|0\rangle_{n+1}...,\nonumber\\
|k_1,k_2,scatt\rangle_0^{\alpha}&=&Z^{-1}_{scatt}(k_1,k_2)\sum_{m=-\infty}^{\infty}
\sum_{n=m+1}^{\infty}
\varepsilon_{\alpha\beta\gamma}a^{scatt}(m,n;k_1,k_2)...|1\rangle_m^{\beta}...|1\rangle_n^{\gamma}
...,\nonumber\\
|k,bound\rangle_0^{\alpha}&=&Z^{-1}_{bound}(k)\sum_{m=-\infty}^{\infty}\sum_{n=m+1}^{\infty}
\varepsilon_{\alpha\beta\gamma}a^{bound}(m,n;k)
...|1\rangle_m^{\beta}...|1\rangle_n^{\gamma}...,
\end{eqnarray}
where
\begin{eqnarray}
a^{scatt}(m,n;k_1,k_2)&=&C_{12}{\rm
e}^{i(k_1m+k_2n)}-C_{21}{\rm e}^{i(k_2m+k_1n)},\nonumber\\
a^{bound}(m,n;k)&=&{\rm e}^{iu(m+n)+v(m-n)},\quad
u=\frac{k}{2}+(1-\frac{\Delta_1}{|\Delta_1|})\frac{\pi}{2}.
\end{eqnarray}
Here $C_{ab}=\cos\frac{k_a+k_b}{2}-\Delta_1{\rm
e}^{i\frac{k_a-k_b}{2}}$, $\Delta_1=5/4-J_{\|}/J_{c}$, $v>0$ and
\begin{equation}
\cos\frac{k}{2} =|\Delta_1|{\rm e}^{-v}.
\end{equation}

The normalization factors,
\begin{eqnarray}
Z_{scatt}(k_1,k_2)&=&\sqrt{2}N\sqrt{\cos^2\frac{k_1+k_2}{2}-2\Delta_1\cos\frac{k_1+k_2}{2}
\cos\frac{k_1-k_2}{2}+\Delta_1^2},
\nonumber\\
Z_{bound}(k)&=&\sqrt{\frac{N\cos^2\frac{k}{2}}{\Delta_1^2-\cos^2\frac{k}{2}}},
\end{eqnarray}
depend on $N$ the number of rungs.

The corresponding dispersion laws are the following:
\begin{eqnarray}
E^{magn}(k)&=&J_{\bot}-\frac{3}{2}J_c+J_c\cos k,\\
E^{scatt}(k_1,k_2)&=&2J_{\bot}-3J_c+J_c(\cos
k_1+\cos k_2),\\
E^{bound}(k)&=&2J_{\bot}+(\Delta_1-3)J_c+\frac{J_c}{\Delta_1}\cos^2\frac{k}{2}.
\end{eqnarray}

As it follows from (25) the one-magnon gap $E^{magn}_{gap}$ and the one-magnon zone width
$\Delta E^{magn}$ are given by the following formulas:
\begin{equation}
E^{magn}_{gap}=J_{\bot}-\frac{3}{2}J_c-|J_c|,\quad \Delta E^{magn}=2|J_c|.
\end{equation}

\section{First order DSF for ${\hat H}^{s-r}+{\hat H}^{asymm}$}

From (4) and (20) follows that,
\begin{eqnarray}
H^{asymm}_{n,n+1}\varepsilon_{\alpha\beta\gamma}|1\rangle_n^{\beta}|1\rangle_{n+1}^{\gamma}&=&
-i(J^{asymm}_{\|}-J^{asymm}_{frust})|1\rangle_n^{\alpha}|0\rangle_{n+1}\nonumber\\
&+&i(J^{asymm}_{\|}+J^{asymm}_{frust})|0\rangle_n|1\rangle_{n+1}^{\alpha},
\end{eqnarray}
so,
\begin{eqnarray}
^{\alpha}_0\langle q,magn|\hat
H^{asymm}|k_1,k_2,scatt\rangle_0^{\beta}&=&
\frac{2\sqrt{2}J^{asymm}(q)\cos\frac{q}{2}\sin\frac{k_1-k_2}{2}\delta_{k_1+k_2\,q}
\delta_{\alpha\beta}}
{\sqrt{N(\cos^2\frac{q}{2}-2\Delta_1\cos\frac{q}{2}\cos\frac{k_1-k_2}{2}+
\Delta_1^2})},\nonumber\\
^{\alpha}_0\langle q,magn|\hat
H^{asymm}|k,bound\rangle_0^{\beta}&=&
-2i\bar J^{asymm}(q)\frac{\sqrt{\Delta_1^2-\cos^2\frac{q}{2}}}{\Delta_1}\delta_{kq}
\delta_{\alpha\beta},
\end{eqnarray}
where
\begin{equation}
J^{asymm}(q)=J_{frust}^{asymm}\cos\frac{q}{2}-iJ_{\|}^{asymm}\sin\frac{q}{2}.
\end{equation}

Considering ${\hat H}^{asymm}$ as a small perturbation we may obtain the corresponding
corrections for the one- and two-magnon states.
In the simplest case when the one-magnon mode does not intersect the
two-magnon sector all the first order corrections to one- and
two-magnon dispersions vanish. First order one-magnon contributions
to the $S=1$ two-magnon states are the following:
\begin{eqnarray}
|k_1,k_2,scatt\rangle_1&=&\frac{2\sqrt{2}J^{asymm}(k_1+k_2)\cos\frac{k_1+k_2}{2}
\sin\frac{k_1-k_2}{2}}
{\sqrt{N(\cos^2\frac{k_1+k_2}{2}-2\Delta_1\cos\frac{k_1+k_2}{2}\cos\frac{k_1-k_2}{2}+
\Delta_1^2)}}\nonumber\\
&\cdot&\frac{1}{(E^{scatt}(k_1,k_2)-E^{magn}(k_1+k_2))}
|k_1+k_2,magn\rangle_0,\nonumber\\
|k,bound\rangle_1&=&-2i\frac{\sqrt{\Delta_1^2-\cos^2
\frac{k}{2}}
\bar J^{asymm}(k)}
{\Delta_1
(E^{bound}(k)-E^{magn}(k))}|k,magn\rangle_0.
\end{eqnarray}

We use the following expression for the zero-temperature dynamical structure
factor \cite{1},\cite{5},\cite{6},
\begin{equation}
S_{\alpha\beta}({\bf
q},\omega)=\lim_{N\rightarrow\infty}\frac{1}{N}\sum_{\mu}\langle0|\hat{\bf
S}^{\alpha}({\bf q})|\mu\rangle\langle\mu|\hat{\bf S}^{\beta}(-{\bf
q})|0\rangle\delta(\omega-E_{\mu}),
\end{equation}
where $\hat{\bf S}({\bf q})$ is the Fourier transformation of spin associated with the two
dimensional vector ${\bf q}=(q,q_{rung})$. Here $q$ and $q_{rung}$ are the corresponding leg and
rung components. Since the latter has only two possible values $0$ and $\pi$ we
may study them separately,
\begin{equation}
\hat{\bf S}(q,0)=\sum_n{\rm e}^{-iqn}({\bf S}_{1,n}+{\bf
S}_{2,n}),\quad \hat{\bf S}(q,\pi)=\sum_n{\rm e}^{-iqn}({\bf
S}_{1,n}-{\bf S}_{2,n}).
\end{equation}

According to the following two formulas,
\begin{equation}
{[}\hat Q,\hat{\bf S}(q,0){]}=0,\quad
\{\hat Q,\hat{\bf S}(q,\pi)\}=\hat{\bf S}(q,\pi),
\end{equation}
we may reduce the matrix elements in (33)
\begin{equation}
\langle\mu|\hat{\bf S}(q,0)|0\rangle=0,\quad
\langle\mu|\hat{\bf S}(q,\pi)|0\rangle=
\sum_{\nu\in{\cal H}^1}\langle\mu|\nu\rangle\langle\nu|
\hat{\bf S}(q,\pi)|0\rangle,
\end{equation}
so $S_{\alpha\beta}(q,0,\omega)=0$. For calculation of $S_{\alpha\beta}(q,\pi,\omega)$ let us
notice that from (20) and (21) follows that
\begin{equation}
\langle 0|\hat{\bf S}^{\alpha}(q,\pi)|k,magn\rangle^{\beta}_0=\sqrt{N}\delta_{\alpha\beta}\delta_{kq},
\end{equation}
so, the DSF has purely diagonal form,
$S_{\alpha\beta}(q,\pi,\omega)=\delta_{\alpha\beta}S(q,\pi,\omega)$, where
\begin{equation}
S(q,\pi,\omega)=\sum_{\mu}|\langle
\mu|q,magn\rangle_0^3|^2\delta(\omega-E_{\mu}).
\end{equation}

The unperturbed DSF corresponding only to ${\hat H}^{s-r}$ consists on a single one-magnon coherent
peak
\begin{equation}
S^{(0)}(q,\pi,\omega)=\delta(\omega-E^{magn}(q)).
\end{equation}

In the first order with respect to the asymmetry we have to take into account only the two-magnon
contributions. Using the substitution
$2\pi\sum_k\rightarrow N\int_{-\pi}^{\pi}dk$ we obtain the
following formula:
\begin{equation}
S^{(1)}(q,\pi,\omega)=A_{bound}(q)\delta(\omega-E^{bound}(q))+A_{scatt}(q,\omega),
\end{equation}
where
\begin{eqnarray}
A_{bound}(q)&=&\frac{4|J^{asymm}(q)|^2(\Delta_1^2-\cos^2
\frac{q}{2})}
{\Delta_1^2
(E^{magn}(q)-E^{bound}(q))^2},\\
A_{scatt}(q,\omega)&=&\frac{4|J^{asymm}(q)|^2(\cos^2\frac{q}{2}-x^2(\omega))\Theta (1-x^2(\omega))}
{\pi(\cos^2\frac{q}{2}-2\Delta_1x(\omega)+
\Delta_1^2)(\omega-E^{magn}(q))^2}.
\end{eqnarray}
Here $\Theta(x)$ is the step function and
$x(\omega)=(\omega-2J_{\bot}+3J_c)/(2J_c)$.

The first term in (40) corresponds to the second coherent peak carried from the sector
${\cal H}^{even}$.

The formulas (41) and (42) will be correct only when the energy of the one-magnon mode is smaller
than the energy of the bound state and the lower bound of the two-magnon continuum. Contrary
due to the asymmetry mixing between ${\cal H}^1$ and ${\cal H}^2$ any
intersection of the one- and two-magnon scattering sectors will result to magnon decay
\cite{10}. In order to avoid this possibility we shall obtain the "non-intersection" condition.

According to (25) and (26),
\begin{equation}
2J_{\bot}-3J_c-2|J_c|\cos\frac{k_1+k_2}{2}\leq E^{scatt}(k_1,k_2)\leq
2J_{\bot}-3J_c+2|J_c|\cos\frac{k_1+k_2}{2},
\end{equation}
and the condition
$E^{magn}(k_1+k_2)<E^{scatt}(k_1,k_2)$ reduces to the following form,
$J_c(\cos k/2+|J_c|/(2J_c))^2<J_{\bot}/2$.
This inequality will be automatically satisfied for $J_c<0$, while for $J_c>0$  it
results to $2J_{\bot}>9J_c$ or using (28) to an equivalent form,
\begin{equation}
E^{magn}_{gap}>\Delta E^{magn}.
\end{equation}

The last formula has a clear interpretation. Really $E^{magn}_{gap}-\Delta E^{magn}$ measures
the difference between the one- and two- magnon sectors. When it is satisfied these sectors do not
intersect a magnon decay is impossible and the formula (42) is correct.

\section{Overview of experimental data for CuHpCl}

As it was suggested in \cite{5} the compound CuHpCl corresponds to the case
$J^{asymm}_{\|}=0$, $J^{asymm}_{frust}=J_{frust}$. In other words it may be described by the
Hamiltonian
\begin{equation}
H^{\rm CuHpCl}_{n,n+1}=H^{rung}_{n,n+1}+H^{leg}_{n,n+1}+H^{diag}_{n,n+1}+H^{cyc}_{n,n+1},
\end{equation}
where the terms $H^{rung}_{n,n+1}$, $H^{leg}_{n,n+1}$ and $H^{cyc}_{n,n+1}$ are given by (3) and
\begin{equation}
H^{diag}_{n,n+1}=J_{diag}{\bf S}_{1,n}{\bf S}_{2,n+1}.
\end{equation}
Here $J_{diag}=J_{frust}+J_{frust}^{asymm}=2J_{frust}$.

If one suggests that the state (10) is the exact ground state then the condition
(8) reduces to
\begin{equation}
2J_{diag}=J_{\|}-\frac{1}{2}J_c.
\end{equation}

Under this condition (however not proved experimentally!) it is possible to estimate the
parameters $J_{\bot}$ and $J_c$ from the formulas (25), (28) and experimental data.
As it was presented in \cite{11} $E_{gap}^{magn}\approx10.8\,{\rm K}$,
corresponds to $k=\pi$, however as it was shown in \cite{12} by $k=0$ ESR measurements
$E_{gap}+2\Delta E^{magn}=13.1\,{\rm K}$. This data agrees with the neutron scattering experiments
\cite{5},\cite{6}. From (25) and (28) follows that
$J_c\approx1.2\,{\rm K}$ and $J_{\bot}\approx13.8\,{\rm K}$.

Unfortunately any quantitative interpretation fails for the neutron scattering data obtained in
\cite{6}. Really all the scans
presented here correspond to the scattering with $q_{rung}=0$. However as it was shown in the
previous section $S_{\alpha\beta}(q,0,\omega)=0$. Therefore appearance of the scattering peaks
in Figs 9 and 10(a) of the Ref. 6 may be explained only by a deviation of the initial state of
the ladder from the singlet-rung vacuum (10). The strength of this deviation may be estimated only
by comparison the data presented in \cite{6} with the same one related to the scattering
with $q_{rung}=\pi$. However the latter is not yet obtained.

In despite of the quantitative disagreement at $q_{rung}=0$ our argumentation qualitatively
confirms the appearance of the second coherent peak in the structure factor.

\section{Summary}
In this paper we have demonstrated the principal difference between the excitation spectrums of
symmetric and asymmetric spin ladders. For the symmetric one the Hilbert space splits on
two invariant subspaces ${\cal H}^{even}$ and ${\cal H}^{odd}$. In this case only the
sector ${\cal H}^{odd}$ gives a nonzero contribution to the dynamical structure factor.
However the picture is quite different for an asymmetric spin ladder. The asymmetry term mixes
both the subspaces and the
two-magnon bound state from ${\cal H}^{even}$ contribute to the DSF resulting to the appearance
of the second coherent peak.

As an illustration we have obtained the first order DSF for the special model of asymmetric
spin-ladder with exact singlet rung ground state.
The suggested model was applied to the probably asymmetric spin-ladder compound CuHpCl
for which the existence of the second coherent peak was observed experimentally. Despite the
full agreement between our special model and the experimental data was not confirmed
some of the interaction constants were estimated from the inelastic neutron scattering
and ESR data.

The authors are very grateful to S.~V.~Maleev for helpful discussions.

\end{document}